\documentclass[prd,aps,floatfix,nofootinbib,11 pt]{revtex4}
\usepackage{amssymb}
\usepackage{amsmath,graphicx,color,epsfig}

\input{tcilatex}

\begin{document}

\title{Quantum memory in non-inertial frames}
\author{M. Ramzan\thanks{%
mramzan@phys.qau.edu.pk} and M. K. Khan}

\address{Department of Physics Quaid-i-Azam University \\
Islamabad 45320, Pakistan}


\begin{abstract}
We study the effect of quantum memory in non-inertial frames under the
influence of amplitude damping, depolarizing, phase flip and bit-phase flip
channels. It is shown that the entanglement of initial state is heavily
influenced by quantum correlations. It is seen that quantum memory
compensates the loss of entanglement caused by the Unruh effect. It is
interesting to note that the sudden death of entanglement disappears for any
acceleration for higher values of quantum memory. Therefore, it is possible
to avoid ESD in non-inertial frames due to the presence of quantum memory.
Furthermore, the degree of entanglement is enhanced as we increase the
degree of memory and it maximizes for maximum correlations.\newline
\end{abstract}

\maketitle

\address{Department of Physics Quaid-i-Azam University \\
Islamabad 45320, Pakistan} 


\vspace*{1.0cm}

\vspace*{1.0cm}



\section{Introduction}

Quantum entanglement being a fundamental potential resource for
communication is one of the key quantitative notions of the intriguing field
of quantum information and quantum computation. Entanglement behavior in
non-inertial frames was first considered by Alsing et al. [1]. They studied
the fidelity of teleportation between relative accelerated partners.
Afterwards, a number of authors have focused on the topic of the quantum
information in a relativistic setup [2-8]. Schuller et al. [9] have studied
how the Unruh effect changes the degree of quantum entanglement. However,
all the investigations in non-inertial frames were confined to the studies
of quantum information in an isolated system except [10, 11], where the
authors have analyzed the decoherence effects in non-inertial frames.

Since in a realistic quantum system, the interaction between the quantum
system and the surrounding environment is inevitable, it is therefore,
important to study the effect of environmental influences on the
entanglement dynamics of a system. Quantum channels can be implemented by
suitable quantum devices consisting of intrinsic degrees of freedom
associated with the environment and acting on the system via particular
interactions between the system and the environment. The assumption that
noise is uncorrelated between successive uses of a channel is not realistic.
Hence memory effects need to be taken into account. Quantum channels with
memory [12-14] are the natural theoretical framework for the study of any
noisy quantum communication system where correlation time is longer than the
time between consecutive uses of the channel. A more general model of a
quantum channel with memory was introduced by Bowen and Mancini [15] and
also studied by Kretschmann and Werner [16].

In this paper, we investigate the effect of quantum memory on the
entanglement of Dirac fields in non-inertial frames. We consider the
amplitude damping, depolarizing, phase flip and bit-phase flip channels. We
assume that two observers, Alice and Rob, share an entangled initial state
at the same point in flat Minkowski spacetime. After that Alice stays
stationary while Rob moves with uniform acceleration. We assume that Alice
and Rob have detectors that are sensitive only to their respective modes and
both the observers share the following maximally entangled initial state%
\begin{equation}
|\Psi \rangle _{AR}=\frac{1}{\sqrt{2}}\left( |0\rangle _{A}|0\rangle
_{R}+|1\rangle _{A}|1\rangle _{R}\right)
\end{equation}%
where the two modes of Minkowski spacetime that correspond to Alice and Rob
are $|\eta \rangle _{A}$ and $|\eta \rangle _{R}$ respectively. From the
perspective of Rob, the Minkowski vacuum is found to be a two-mode squeezed
state [4]%
\begin{equation}
|0\rangle _{M}=\cos r|0\rangle _{I}|0\rangle _{II}+\sin r|1\rangle
_{I}|1\rangle _{II}
\end{equation}%
where $\cos r=(e^{-2\pi \omega c/a}+1)^{-1/2}.$ The constants $\omega ,$ $c$
and $a$, in the exponential stand for Dirac particle's frequency, light's
speed in vacuum and Rob's acceleration respectively. The subscripts $I$ and $%
II$ of the kets represent the Rindler modes in region $I$ and $II$,
respectively, in the Rindler spacetime diagram (see Ref. [10], Fig. (1)).
The excited state in Minkowski spacetime is related to Rindler modes as
follows [4]%
\begin{equation}
|1\rangle _{M}=|1\rangle _{I}|0\rangle _{II}
\end{equation}%
Using equations (2) and (3), equation (1) can be written in terms of
Minkowski modes for Alice and Rindler modes for Rob as
\begin{equation}
|\Psi \rangle _{A,I,II}=\frac{1}{\sqrt{2}}\left( \cos r|0\rangle
_{A}|0\rangle _{I}|0\rangle _{II}|+\sin r|0\rangle _{A}|1\rangle
_{I}|1\rangle _{II}+|1\rangle _{A}|1\rangle _{I}|0\rangle _{II}\right)
\end{equation}%
Since Rob is causally disconnected from region $II$, therefore, by tracing
over all the modes in region $II$ leaves the following mixed density matrix
between Alice and Rob%
\begin{equation}
\rho _{A,I}=\frac{1}{2}\left( \cos ^{2}r|00\rangle \left\langle
00\right\vert +\cos r(|00\rangle \left\langle 11\right\vert +|11\rangle
\left\langle 00\right\vert )+\sin ^{2}r|01\rangle \left\langle 01\right\vert
+|11\rangle \left\langle 11\right\vert \right)
\end{equation}%
We study the entanglement of Dirac fields in non-inertial frames influenced
by different memory channels, such as amplitude damping, depolarizing, phase
flip and bit-phase-flip channels, parameterized by decoherence parameter $p$%
\ and memory parameter $%
\mu
$. Here $p\in \lbrack 0,1]$ and $\mu \in \lbrack 0,1]$ represent the lower
and upper limits of decoherence parameter and memory parameter respectively.
It is seen that depolarizing channel influences the entanglement of Dirac
fields heavily in comparison to the amplitude damping and flipping channels.
Therefore, the entanglement of Dirac fields is strongly dependent on degree
of correlations of the noisy channels.

\section{Entanglement in a correlated environment}

Since noise is a major hurdle while transmitting quantum information from
one party to other through classical and quantum channels. This noise causes
a distortion of the information sent through the channel. If multiple uses
of a channel are not correlated, there is no advantage in using entangled
states. Correlated noise, also referred as memory in the literature, acts on
consecutive uses of the channel. However, in general, one may want to encode
classical data into entangled strings or consecutive uses of the channel may
be correlated to each other. Hence, we are dealing with a strongly
correlated quantum system, the correlation of which results from the memory
of the channel itself. In\ Ref. [12], a Pauli channel with partial memory
was studied. The action of such a channel on two consecutive qubits is given
in Kraus operator form as%
\begin{equation}
A_{ij}=\sqrt{p_{i}[(1-\mu )p\alpha _{j}+\mu \delta _{ij}]}\sigma _{i}\otimes
\sigma _{j}
\end{equation}%
where $\sigma _{i}$ ($\sigma _{j})$ are usual Pauli matrices, $p_{i}$ ($%
p_{j} $) represent the quantum noise and indices $i$ and $j$ runs from $0$
to $3.$ The above expression means that with probability $\mu $ the channel
acts on the second qubit with the same error operator as on the first qubit,
and with probability $(1-\mu )$ it acts on the second qubit independently.
Physically the parameter $%
\mu
$ is determined by the relaxation time of the channel when a qubit passes
through it. In order to remove correlations, one can wait until the channel
has relaxed to its original state before sending the next qubit, however
this lowers the rate of information transfer. Thus it is necessary to
consider the performance of the channel for arbitrary values of $%
\mu
$ to reach a compromise between various factors which determine the final
rate of information transfer.\ Thus in passing through the channel any two
consecutive qubits undergo random independent (uncorrelated) errors with
probability ($1-%
\mu
)$ and identical (correlated) errors with probability $%
\mu
$. This should be the case if the channel has a memory depending on its
relaxation time and if we stream the qubits through it. A detailed list of
single qubit Kraus operators for different quantum channels with
uncorrelated noise is given in table 1. The action of such a channel if $n$
qubits are streamed through it, can be described in operator sum
representation as [17]
\begin{equation}
\rho _{f}=\sum\limits_{k_{1,}....,.k_{n}=0}^{n-1}(A_{k_{n}}\otimes
.....A_{k_{1}})\rho _{in}(A_{k_{1}}^{\dagger }\otimes
.....A_{k_{n}}^{\dagger })
\end{equation}%
where $\rho _{in}$ represents the initial density matrix for quantum state
and $A_{k_{n}}$\ are the Kraus operators expressed in equation (6). The
Kraus operators satisfy the completeness relation
\begin{equation}
\sum\limits_{k_{n}=0}^{n-1}A_{k_{n}}^{\dagger }A_{k_{n}}=1
\end{equation}%
However, the Kraus operators for a quantum amplitude damping channel with
correlated noise are given by Yeo and Skeen [13] as given as%
\begin{equation}
A_{00}^{c}=\left[
\begin{array}{llll}
\cos \chi & 0 & 0 & 0 \\
0 & 1 & 0 & 0 \\
0 & 0 & 1 & 0 \\
0 & 0 & 0 & 1%
\end{array}%
\right] ,\ \ \ A_{11}^{c}=\left[
\begin{array}{llll}
0 & 0 & 0 & 0 \\
0 & 0 & 0 & 0 \\
0 & 0 & 0 & 0 \\
\sin \chi & 0 & 0 & 0%
\end{array}%
\right]
\end{equation}%
where, $0\leq \chi \leq \pi /2$ and is related to the quantum noise
parameter as
\begin{equation}
\sin \chi =\sqrt{p}
\end{equation}%
It is clear that $A_{00}^{c}$ cannot be written as a tensor product of two
two-by-two matrices. This gives rise to the typical spooky action of the
channel: $\left\vert 01\right\rangle $ and $\left\vert 10\right\rangle $,
and any linear combination of them, and $\left\vert 11\right\rangle $ will
go through the channel undisturbed, but not $\left\vert 00\right\rangle .$
The action of this non-unital channel is given by
\begin{equation}
\pi \rightarrow \rho =\Phi (\pi )=(1-\mu
)\sum\limits_{i,j=0}^{1}A_{ij}^{u}\pi A_{ij}^{u\dagger }+\mu
\sum\limits_{k=0}^{1}A_{kk}^{c}\pi A_{kk}^{c\dagger }
\end{equation}%
The action of the super-operators provides a way of describing the evolution
of quantum states in a noisy environment. In our scheme, the Kraus operators
are of the dimension $2^{2}$. They are constructed from single qubit Kraus
operators by taking their tensor product over all $n^{2}$ combinations
\begin{equation}
A_{k}=\underset{k_{n}}{\otimes }A_{k_{n}}
\end{equation}%
where $n$ is the number of Kraus operator for a single qubit channel. The
final state of the system after the action of the channel can be obtained as
\begin{equation}
\rho _{f}=\Phi _{p,\mu }(\rho _{A,I})
\end{equation}%
where $\Phi _{p,\mu }$ is the super-operator realizing the quantum channel
parametrized by real numbers $p$ and $\mu $. The density matrix after the
action of different memory channels are obtained by using equations (5-13).
It is important to mention here that we consider the case that both the
qubits are coupled to the time correlated channel with memory. The final
density matrix after the action of amplitude damping channel is given by%
\begin{eqnarray}
\rho _{f}^{\text{AD}} &=&\left(
\begin{array}{cccc}
\left.
\begin{array}{c}
\frac{1}{2}(-(-1+\mu )p(1+p) \\
-(-1+p)\cos ^{2}r)%
\end{array}%
\right. & 0 & 0 & \left.
\begin{array}{c}
\frac{1}{2}(1-p+\mu (-1 \\
+\sqrt{(1-p)}+p))\cos r%
\end{array}%
\right. \\
0 & \left.
\begin{array}{c}
\frac{1}{2}(1+(-1+\mu )p^{2} \\
+(-1+p-\mu p)\cos ^{2}r)%
\end{array}%
\right. & 0 & 0 \\
0 & 0 & \left.
\begin{array}{c}
\frac{1}{2}(-1+\mu ) \\
(-1+p)p%
\end{array}%
\right. & 0 \\
\left.
\begin{array}{c}
\frac{1}{2}(1-p+\mu (-1 \\
+\sqrt{(1-p)}+p))\cos r%
\end{array}%
\right. & 0 & 0 & \left.
\begin{array}{c}
\frac{1}{2}(1-(-1+\mu ) \\
(-2+p)p+\mu p\cos ^{2}r)%
\end{array}%
\right.%
\end{array}%
\right)  \notag \\
&&
\end{eqnarray}%
The final density matrix after the action of depolarizing channel is given by

\begin{eqnarray}
\rho _{f}^{\text{Dep}} &=&\left(
\begin{array}{cccc}
\left.
\begin{array}{c}
\frac{1}{8}\cos ^{2}r(2+p(-3 \\
-2\mu (-2+p)+2p) \\
-(-2+p)(1+(-1 \\
+\mu )p)\cos ^{2}r)%
\end{array}%
\right. & 0 & 0 & \left.
\begin{array}{c}
\frac{1}{16}\cos r(4+(-2 \\
+\mu )p+(4+p(-10 \\
+\mu (11-6p)+6p)) \\
\cos ^{2}r)%
\end{array}%
\right. \\
0 & \left.
\begin{array}{c}
\frac{1}{16}(4+(-2+\mu )p \\
+2(-2+p)(1+(-1 \\
+\mu )p)\cos ^{2}r)\sin ^{2}r%
\end{array}%
\right. & -\frac{1}{16}\mu p\cos r\sin ^{2}r & 0 \\
0 & 0 & 0 & 0 \\
\left.
\begin{array}{c}
\frac{1}{8}\cos r(2+p(-3 \\
-2\mu (-2+p)+2p) \\
-(-2+p)(1+(-1 \\
+\mu )p)\cos ^{2}r)%
\end{array}%
\right. & 0 & 0 & \left.
\begin{array}{c}
\frac{1}{16}(4+(-2+\mu )p+ \\
(4+p(-10+\mu (11 \\
-6p)+6p))\cos ^{2}r)%
\end{array}%
\right.%
\end{array}%
\right)  \notag \\
&&
\end{eqnarray}%
The final density matrix after the action of bit-phase flip channel is given
by

\begin{eqnarray}
\rho _{f}^{\text{Bpf}} &=&\left(
\begin{array}{cccc}
\left.
\begin{array}{c}
\frac{1}{4}(1+p(-1-2\mu (-1+p) \\
+2p)-(-1+p)(1 \\
+2(-1+\mu )p)\cos (2r))%
\end{array}%
\right. & 0 & 0 & \left.
\begin{array}{c}
\frac{1}{16}\cos r(4+(-2+\mu )p \\
+(4+p(-10+\mu (11 \\
-6p)+6p))\cos ^{2}r)%
\end{array}%
\right. \\
0 & \left.
\begin{array}{c}
\frac{1}{16}(4+(-2+\mu )p+ \\
2(-2+p)(1+(-1 \\
+\mu )p)\cos ^{2}r)\sin ^{2}r%
\end{array}%
\right. & \left.
\begin{array}{c}
-\frac{1}{16}\mu p \\
\cos r\sin ^{2}r%
\end{array}%
\right. & 0 \\
0 & 0 & 0 & 0 \\
\left.
\begin{array}{c}
\frac{1}{8}\cos r(2+p(-3-2\mu \\
(-2+p)+2p)-(-2+p) \\
(1+(-1+\mu )p)\cos ^{2}r)%
\end{array}%
\right. & 0 & 0 & \left.
\begin{array}{c}
\frac{1}{16}(4+(-2+\mu )p \\
+(4+p(-10+\mu (11 \\
-6p)+6p))\cos ^{2}r)%
\end{array}%
\right.%
\end{array}%
\right)  \notag \\
&&
\end{eqnarray}%
The final density matrix after the action of phase flip channel is given by

\begin{equation}
\rho _{f}^{\text{Pf}}=\left(
\begin{array}{cccc}
\frac{1}{2}\cos ^{2}r & 0 & 0 & \left.
\begin{array}{c}
\frac{1}{2}(1+4p(-1+\mu \\
+p-\mu p))\cos r%
\end{array}%
\right. \\
0 & \frac{1}{2}\sin ^{2}r & -\frac{1}{16}\mu p\cos r\sin ^{2}r & 0 \\
0 & 0 & 0 & 0 \\
\left.
\begin{array}{c}
\frac{1}{2}(1+4p(-1+ \\
\mu +p-\mu p))\cos r%
\end{array}%
\right. & 0 & 0 & \frac{1}{2}%
\end{array}%
\right)
\end{equation}%
where the super-scripts AD, Dep, Bpf and Pf correspond to the amplitude
damping, depolarizing, bit-phase flip and phase flip channels respectively.
In order to study the degree of entanglement in the two qubits mixed state
in a correlated noisy environment we calculate the concurrence $C$, which is
given as [18]%
\begin{equation}
C=\max \{0,\sqrt{\lambda _{1}}-\sqrt{\lambda _{2}}-\sqrt{\lambda 3}-\sqrt{%
\lambda _{4}}\}\qquad \lambda _{i}\geqslant \lambda _{i+1}\geqslant 0
\end{equation}%
where $\lambda _{i}$ are the eigenvalues of the matrix $\rho _{f}\tilde{\rho}%
_{f},$ with $\tilde{\rho}_{f}$ being the spin flip matrix of $\rho _{f}$ (as
given in equations (14-17)) and is given as%
\begin{equation}
\tilde{\rho}_{f}=(\sigma _{y}\otimes \sigma _{y})\rho _{f}(\sigma
_{y}\otimes \sigma _{y})
\end{equation}%
where $\sigma _{y}$ is the usual Pauli matrix. The eigen values of $\rho
_{f}^{\text{AD}}\tilde{\rho}_{f}^{\text{AD}}$ are given as under%
\begin{eqnarray}
\lambda _{1,2} &=&\frac{1}{4}(p-\mu p-p^{2}+3\mu p^{2}-2\mu
^{2}p^{2}-p^{3}+2\mu p^{3}-\mu ^{2}p^{3}+p^{4}  \notag \\
&&-2\mu p^{4}+\mu ^{2}p^{4}+2\cos ^{2}r-2\mu \cos ^{2}r+2\mu ^{2}\cos
^{2}r+2\mu \sqrt{1-p}\cos ^{2}r  \notag \\
&&-2\mu ^{2}\sqrt{1-p}\cos ^{2}r-5p\cos ^{2}r+6\mu p\cos ^{2}r-3\mu
^{2}p\cos ^{2}r-2\mu \sqrt{1-p}p\cos ^{2}r  \notag \\
&&+2\mu ^{2}\sqrt{1-p}p\cos ^{2}r+4p^{2}\cos ^{2}r-4\mu p^{2}\cos
^{2}r-p^{3}\cos ^{2}r+2\mu p^{3}\cos ^{2}r  \notag \\
&&-\mu ^{2}p^{3}\cos ^{2}r+\mu p\cos r^{4}-\mu p^{2}\cos r^{4}  \notag \\
&&\pm \frac{1}{\sqrt{2}}\sqrt{\left.
\begin{array}{c}
((-1+p)(-1+p-2\mu (-1+\sqrt{1-p}+p) \\
+\mu ^{2}(-2+2\sqrt{1-p}+p))\cos ^{2}r \\
(4-4p+3\mu p+4p^{2}+13\mu p^{2}-20\mu ^{2}p^{2} \\
-12p^{3}+24\mu p^{3}-12\mu ^{2}p^{3}+8p^{4}-16\mu p^{4}+8\mu ^{2}p^{4} \\
-4(-1-3(-1+\mu )p+(-3+3\mu +\mu ^{2})p^{2} \\
+(-1+\mu )^{2}p^{3})\cos (2r)-\mu (-1+p)p\cos (4r)))))%
\end{array}%
\right. }  \notag \\
\lambda _{3,4} &=&\frac{1}{4}(-1+\mu )(-1+p)p(1+(-1+\mu )p^{2}+(-1+p-\mu
p)\cos ^{2}r)
\end{eqnarray}%
Similarly, the eigen values of $\rho _{f}^{\text{Dep}}\tilde{\rho}_{f}^{%
\text{Dep}}$ are given below%
\begin{eqnarray}
\lambda _{1} &=&\frac{1}{32}\cos ^{2}r(8+2(-8+9\mu )p+(14-19\mu +4\mu
^{2})p^{2}  \notag \\
&&-2(2-3\mu +\mu ^{2})p^{3}+(16+48(-1+\mu )p  \notag \\
&&+2(30-52\mu +23\mu ^{2})p^{2}+(-40+87\mu -47\mu ^{2})p^{3}  \notag \\
&&+12(-1+\mu )^{2}p^{4})\cos ^{2}r+(-2+p)(-4+(14-15\mu )p  \notag \\
&&+(-16+27\mu -11\mu ^{2})p^{2}+6(-1+\mu )^{2}p^{3})\cos r^{4})  \notag \\
\lambda _{2,3,4} &=&0
\end{eqnarray}%
and the eigen values of $\rho _{f}^{\text{Bpf}}\tilde{\rho}_{f}^{\text{Bpf}}$
are obtained as%
\begin{eqnarray}
\lambda _{1,2} &=&\frac{1}{32}(8-27p+30\mu p+55p^{2}-86\mu p^{2}+28\mu
^{2}p^{2}-56p^{3}  \notag \\
&&+112\mu p^{3}-56\mu ^{2}p^{3}+28p^{4}-56\mu p^{4}+28\mu ^{2}p^{4}  \notag
\\
&&+8\cos (2r)-36p\cos (2r)+32\mu p\cos (2r)+68p^{2}\cos (2r)  \notag \\
&&-96\mu p^{2}\cos (2r)+32\mu ^{2}p^{2}\cos (2r)-64p^{3}\cos (2r)  \notag \\
&&+128\mu p^{3}\cos (2r)-64\mu ^{2}p^{3}\cos (2r)+32p^{4}\cos (2r)  \notag \\
&&-64\mu p^{4}\cos (2r)+32\mu ^{2}p^{4}\cos (2r)-p\cos (4r)  \notag \\
&&+2\mu p\cos (4r)+5p^{2}\cos (4r)-10\mu p^{2}\cos (4r)  \notag \\
&&+4\mu ^{2}p^{2}\cos (4r)-8p^{3}\cos (4r)+16\mu p^{3}\cos (4r)  \notag \\
&&-8\mu ^{2}p^{3}\cos (4r)+4p^{4}\cos (4r)-8\mu p^{4}\cos (4r)+4\mu
^{2}p^{4}\cos (4r)  \notag \\
&&\pm 4\sqrt{2}\sqrt{\left.
\begin{array}{c}
((1+2(-1+\mu )p-2(-1+\mu )p^{2})^{2}\cos ^{2}r(4-11p \\
+14\mu p+23p^{2}-38\mu p^{2}+12\mu ^{2}p^{2} \\
-24p^{3}+48\mu p^{3}-24\mu ^{2}p^{3}+12p^{4}-24\mu p^{4} \\
+12\mu ^{2}p^{4}+4(1+(-5+4\mu )p+(3-2\mu )^{2}p^{2} \\
-8(-1+\mu )^{2}p^{3}+4(-1+\mu )^{2}p^{4})\cos (2r) \\
+(-1+p)p((1-2p)^{2}-2\mu (1-2p)^{2}+4\mu ^{2}(-1+p)p)\cos (4r))))%
\end{array}%
\right. }  \notag \\
\lambda _{3,4} &=&\frac{1}{32}(5p-2\mu p+23p^{2}-54\mu p^{2}+28\mu
^{2}p^{2}-56p^{3}+112\mu p^{3}-56\mu ^{2}p^{3}  \notag \\
&&+28p^{4}-56\mu p^{4}+28\mu ^{2}p^{4}-4p\cos (2r)+36p^{2}\cos (2r)-64\mu
p^{2}\cos (2r)  \notag \\
&&+32\mu ^{2}p^{2}\cos (2r)-64p^{3}\cos (2r)+128\mu p^{3}\cos (2r)-64\mu
^{2}p^{3}\cos (2r)  \notag \\
&&+32p^{4}\cos (2r)-64\mu p^{4}\cos (2r)+32\mu ^{2}p^{4}\cos (2r)-p\cos (4r)
\notag \\
&&+2\mu p\cos (4r)+5p^{2}\cos (4r)-10\mu p^{2}\cos (4r)+4\mu ^{2}p^{2}\cos
(4r)  \notag \\
&&-8p^{3}\cos (4r)+16\mu p^{3}\cos (4r)-8\mu ^{2}p^{3}\cos (4r)  \notag \\
&&+4p^{4}\cos (4r)-8\mu p^{4}\cos (4r)+4\mu ^{2}p^{4}\cos (4r)  \notag \\
&&\pm 8\sqrt{2}\sqrt{\left.
\begin{array}{c}
((-1+\mu )^{2}(-1+p)^{3}p^{3}\cos ^{2}r(-5+2\mu -12p+24\mu p \\
-12\mu ^{2}p+12p^{2}-24\mu p^{2}+12\mu ^{2}p^{2} \\
+4(1-4(-1+\mu )^{2}p+4(-1+\mu )^{2}p^{2})\cos (2r) \\
+((1-2p)^{2}-2\mu (1-2p)^{2}+4\mu ^{2}(-1+p)p)\cos (4r))))%
\end{array}%
\right. }
\end{eqnarray}%
The eigen values of $\rho _{f}^{\text{Pf}}\tilde{\rho}_{f}^{\text{Pf}}$ are
given as under%
\begin{eqnarray}
\lambda _{1} &=&(1+2(-1+\mu )p-2(-1+\mu )p^{2})^{2}\cos ^{2}r  \notag \\
\lambda _{2} &=&4(-1+\mu )^{2}(-1+p)^{2}p^{2}\cos ^{2}r  \notag \\
\lambda _{3,4} &=&0
\end{eqnarray}%
The concurrence $C$, is calculated using equation (18). It can be easily
checked from equations (20-23) that the concurrence becomes $\cos r$ if we
set $p=\mu =0,$ which reproduces the results of Ref. [4]. Similarly the
results of Ref [10-11] can be reproduced by setting $\mu =0.$

\section{Discussions}

We have computed the mathematical relations for concurrence in case of
amplitude damping, depolarizing, phase flip and bit-phase flip channels.
However, due to very lengthy mathematical expressions, we have only plotted
it for different parameters such as $p$, $r$, $\mu $\ respectively.

In figures 1 and 2, we plot the concurrence as a function of memory
parameter $\mu $\ for different values of Rob's acceleration, $r$ for
decoherence parameter $p=0.5$\ for amplitude damping, depolarizing channels
respectively. In figures 3 and 4, we plot the concurrence as a function of
memory parameter $\mu $\ with others parameters similar\ as in figures 1 and
2 for bit-phase flip and phase flip channels respectively. It is seen that
quantum memory enhances the entanglement of Dirac fields in contrast to the
effect of decoherence. It is also seen that the depolarizing channel
influences the entanglement more heavily in comparison to the other
channels. The depolarizing channel enhances the degree of entanglement upto
1 even with 50\% decoherence (see figure 2). It is clear from figure 3 that
for lower values of memory parameter, the decoherence dominates and ESD
behaviour can be seen. However for higher values of memory, the ESD can also
be avoided for flipping channels (see figures 3 and 6).

In figure 5, we plot the concurrence as a function of Rob's acceleration $r$%
\ for $p=\mu =0.5$\ for amplitude damping, depolarizing, bit-phase flip and
phase flip channels. It is seen that the entanglement degradation is
controlled due to the presence of quantum memory even at 50\% decoherence.
It is also seen that entanglement is heavily degraded by amplitude damping
channel as compared to other channels. In figure 6, we plot the concurrence
as a function of decoherence parameter $p$ for $\mu =0.5$\ and $r=\pi /6$
for amplitude damping, depolarizing, bit-phase flip and phase flip channels.
It is seen that the maximum entanglement degradation occurs for amplitude
damping channel as compared to other channels. Furthermore, the symmetric
behaviour of flipping channels is seen with maximum degradation for 50\%
decoherence. In order to check the validity of our calculations, we plot in
figure 7, the concurrence as a function of decoherence parameter $p$ for $%
\mu =0$\ and $r=\pi /10$ for amplitude damping, depolarizing, bit-phase flip
and phase flip channels. It is seen that the maximum entanglement
degradation occurs for bit-phase flip channel as compared to other channels.
It is clear from the figure that the graphs of Ref. [10, 11] are reproduced
in the absence of memory.

\section{Conclusions}

We study the effect of quantum memory in non-inertial frames for amplitude
damping, depolarizing, phase flip and bit-phase flip channels. It is shown
that the entanglement of initial state is heavily influenced by quantum
correlations. It is seen that quantum memory compensates the loss of
entanglement generated by the Unruh effect and decoherence. It is
interesting to note that the sudden death of entanglement disappears for any
acceleration for higher degree of memory. Furthermore, the entanglement of
Dirac fields increases as we increase the degree of memory. In conclusion,
we can say that quantum memory can avoid ESD caused by Unruh effect for all
the channels considered.\newline

{\huge Figures captions}\newline
\textbf{Figure 1}. The concurrence is plotted as a function of memory
parameter $\mu $\ for different values of Rob's acceleration, $r=0$ (solid
line), $r=\pi /6$ (dashed line) and $r=\pi /4$ (dotted line) for decoherence
parameter $p=0.5$\ for amplitude damping.\newline
\textbf{Figure 2}. The concurrence is plotted as a function of memory
parameter $\mu $\ for different values of Rob's acceleration, $r=0$ (solid
line), $r=\pi /6$ (dashed line) and $r=\pi /4$ (dotted line) for decoherence
parameter $p=0.5$\ for depolarizing channel.\newline
\textbf{Figure 3}. The concurrence is plotted as a function of memory
parameter $\mu $\ for different values of Rob's acceleration, $r=0$ (solid
line), $r=\pi /6$ (dashed line) and $r=\pi /4$ (dotted line) for decoherence
parameter $p=0.5$\ for bit-phase flip channel.\newline
\textbf{Figure 4}. The concurrence is plotted as a function of memory
parameter $\mu $\ for different values of Rob's acceleration, $r=0$ (solid
line), $r=\pi /6$ (dashed line) and $r=\pi /4$ (dotted line) for decoherence
parameter $p=0.5$\ for phase flip channel.\newline
\textbf{Figure 5}. The concurrence is plotted as a function of Rob's
acceleration $r$\ for $p=\mu =0.5$\ for amplitude damping (solid line),
depolarizing (dashed line), bit-phase flip (dotted line) and phase flip (dot
dashed line) channels.\newline
\textbf{Figure 6}. The concurrence is plotted as a function of decoherence
parameter $p$ for $\mu =0.5$\ and $r=\pi /6$ for amplitude damping (solid
line), depolarizing (dashed line), bit-phase flip (dotted line) and phase
flip (dot dashed line) channels.\newline
\textbf{Figure 7}. The concurrence is plotted as a function of decoherence
parameter $p$ for $\mu =0$\ and $r=\pi /10$ for amplitude damping (solid
line), depolarizing (dashed line), bit-phase flip (dotted line) and phase
flip (dot dashed line) channels.\newline
{\Huge Table Caption}\newline
\textbf{Table 1}. Single qubit Kraus operators for amplitude damping,
depolarizing, bit-phase flip and phase flip channels where $p$ represents
the decoherence parameter.\newline

\begin{figure}[tbp]
\begin{center}
\vspace{-2cm} \includegraphics[scale=0.8]{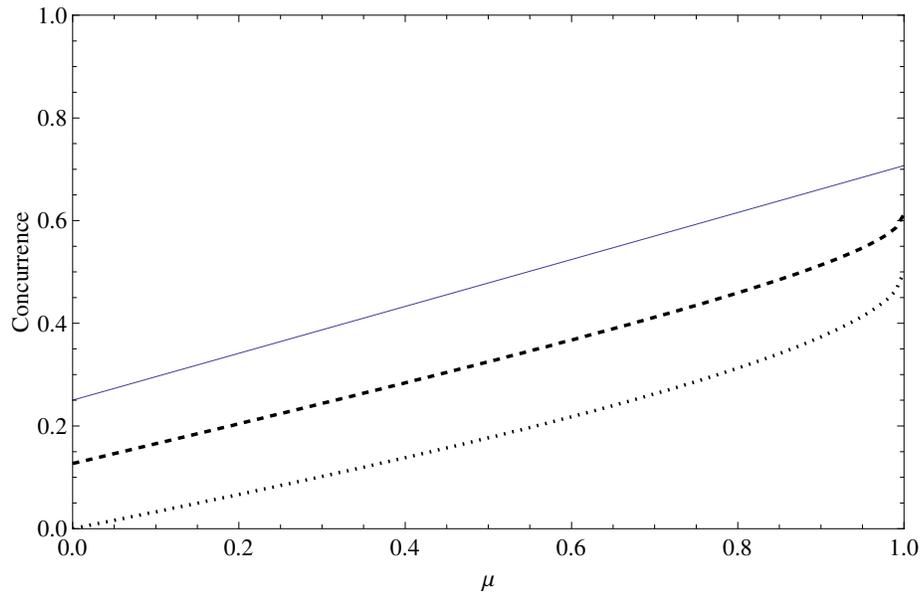} \\[0pt]
\end{center}
\caption{The concurrence is plotted as a function of memory parameter $%
\protect\mu $\ for different values of Rob's acceleration, $r=0$ (solid
line), $r=\protect\pi /6$ (dashed line) and $r=\protect\pi /4$ (dotted line)
for decoherence parameter $p=0.5$\ for amplitude damping.}
\end{figure}

\begin{figure}[tbp]
\begin{center}
\vspace{-2cm} \includegraphics[scale=0.8]{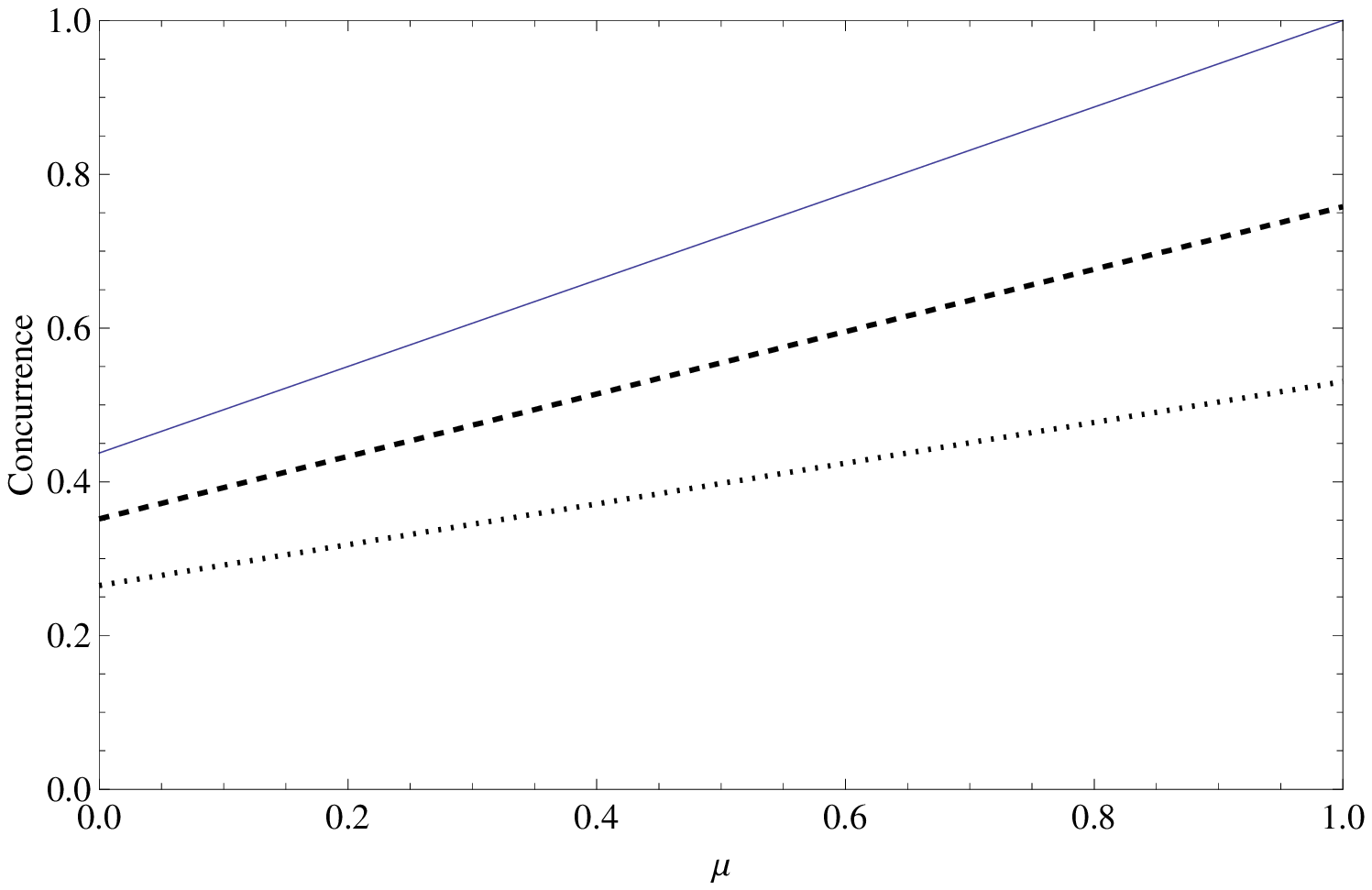} \\[0pt]
\end{center}
\caption{The concurrence is plotted as a function of memory parameter $%
\protect\mu $\ for different values of Rob's acceleration, $r=0$ (solid
line), $r=\protect\pi /6$ (dashed line) and $r=\protect\pi /4$ (dotted line)
for decoherence parameter $p=0.5$\ for depolarizing channel.}
\end{figure}

\begin{figure}[tbp]
\begin{center}
\vspace{-2cm} \includegraphics[scale=0.8]{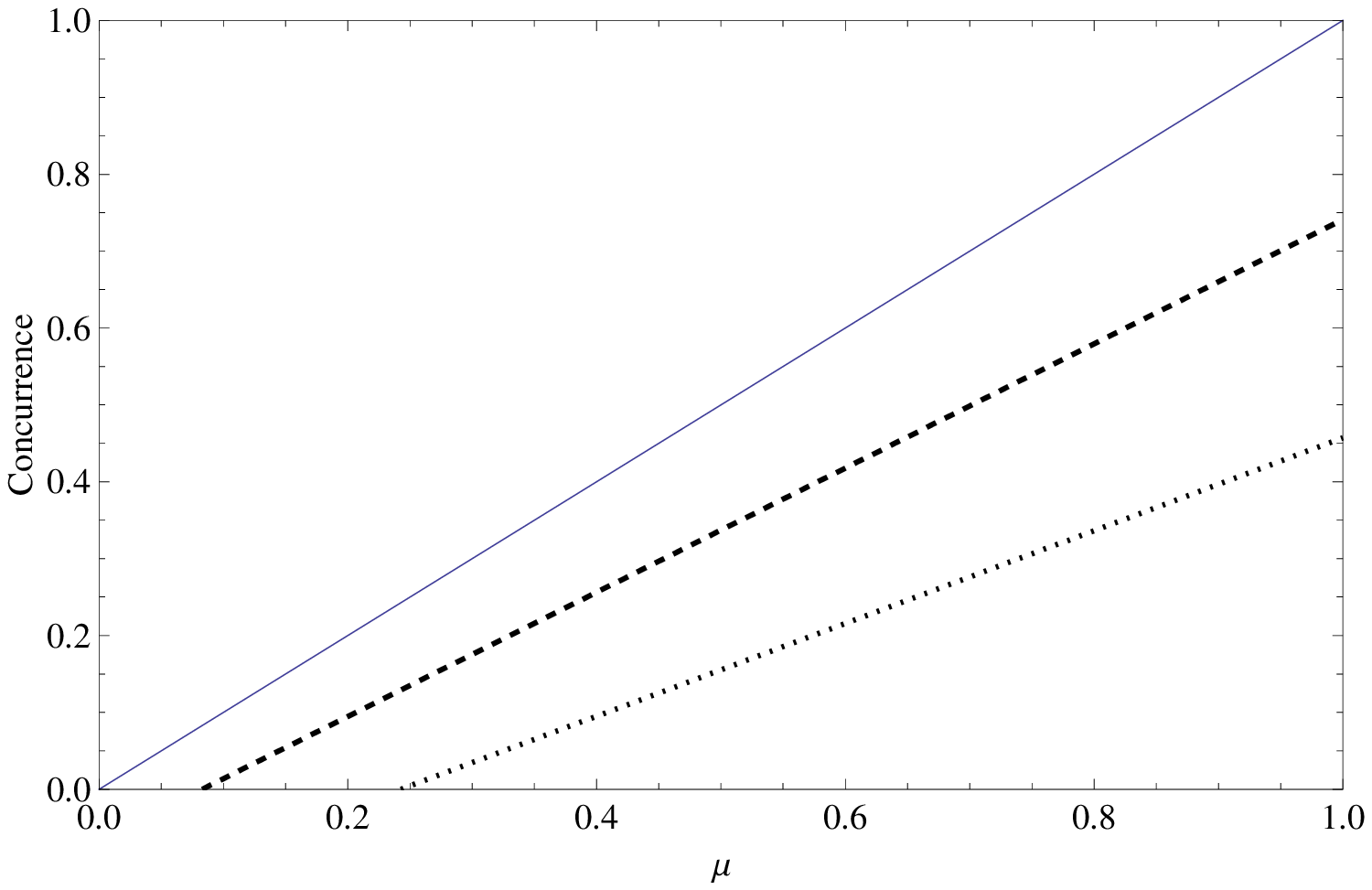} \\[0pt]
\end{center}
\caption{The concurrence is plotted as a function of memory parameter $%
\protect\mu $\ for different values of Rob's acceleration, $r=0$ (solid
line), $r=\protect\pi /6$ (dashed line) and $r=\protect\pi /4$ (dotted line)
for decoherence parameter $p=0.5$\ for bit-phase flip channel.}
\end{figure}

\begin{figure}[tbp]
\begin{center}
\vspace{-2cm} \includegraphics[scale=0.8]{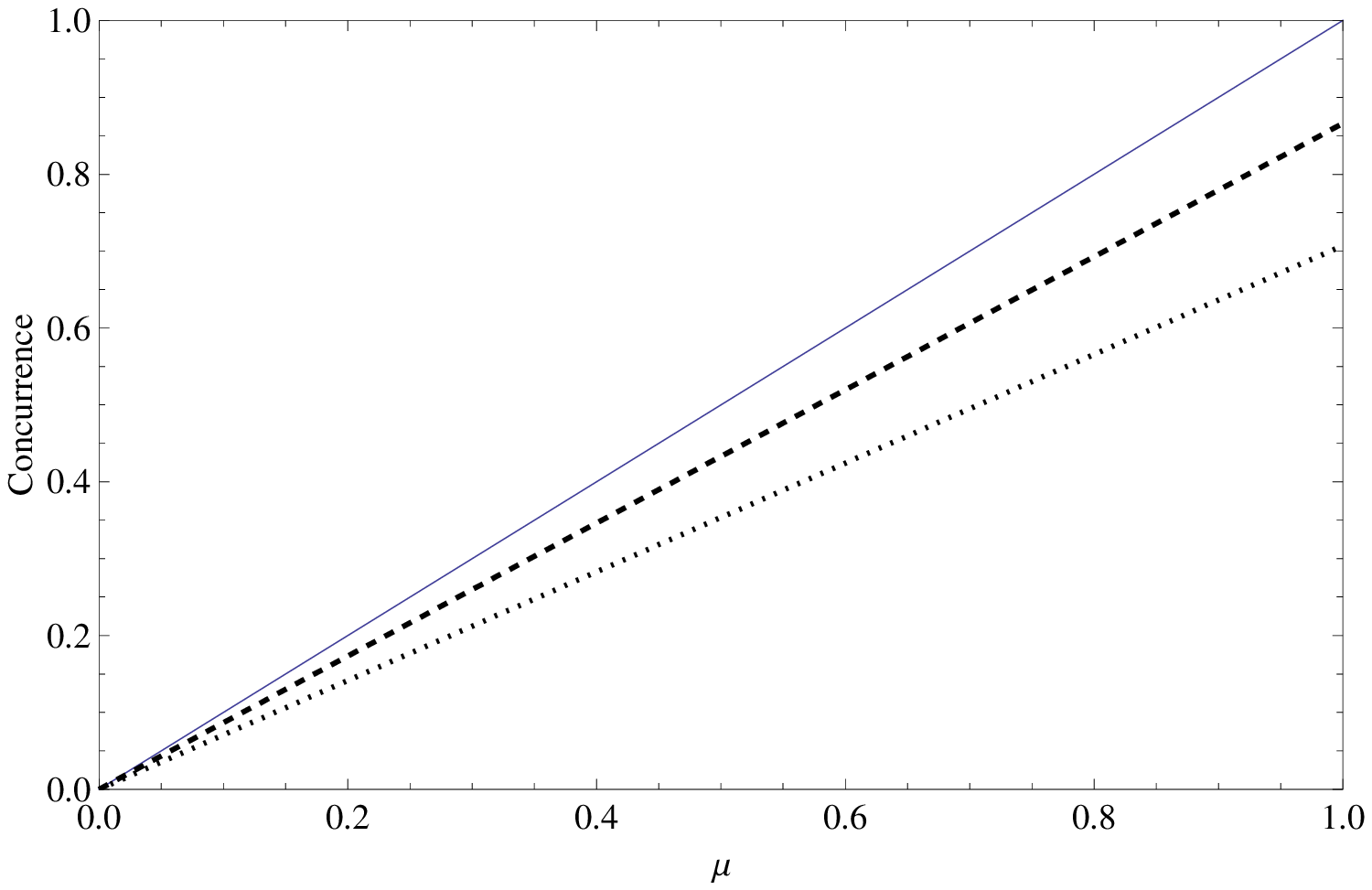} \\[0pt]
\end{center}
\caption{The concurrence is plotted as a function of memory parameter $%
\protect\mu $\ for different values of Rob's acceleration, $r=0$ (solid
line), $r=\protect\pi /6$ (dashed line) and $r=\protect\pi /4$ (dotted line)
for decoherence parameter $p=0.5$\ for phase flip channel.}
\end{figure}

\begin{figure}[tbp]
\begin{center}
\vspace{-2cm} \includegraphics[scale=0.8]{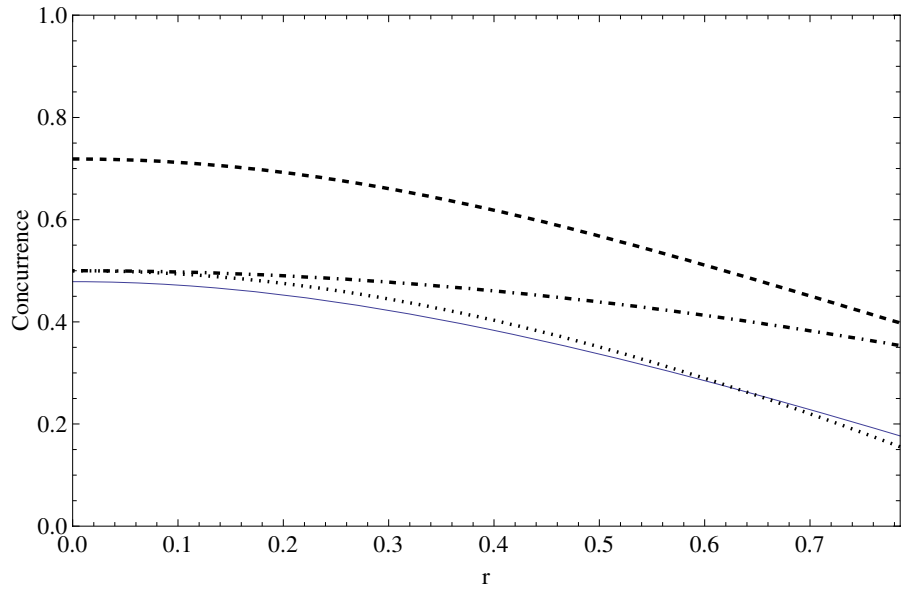} \\[0pt]
\end{center}
\caption{The concurrence is plotted as a function of Rob's acceleration $r$\
for $p=\protect\mu =0.5$\ for amplitude damping (solid line), depolarizing
(dashed line), bit-phase flip (dotted line) and phase flip (dot dashed line)
channels.}
\end{figure}

\begin{figure}[tbp]
\begin{center}
\vspace{-2cm} \includegraphics[scale=0.8]{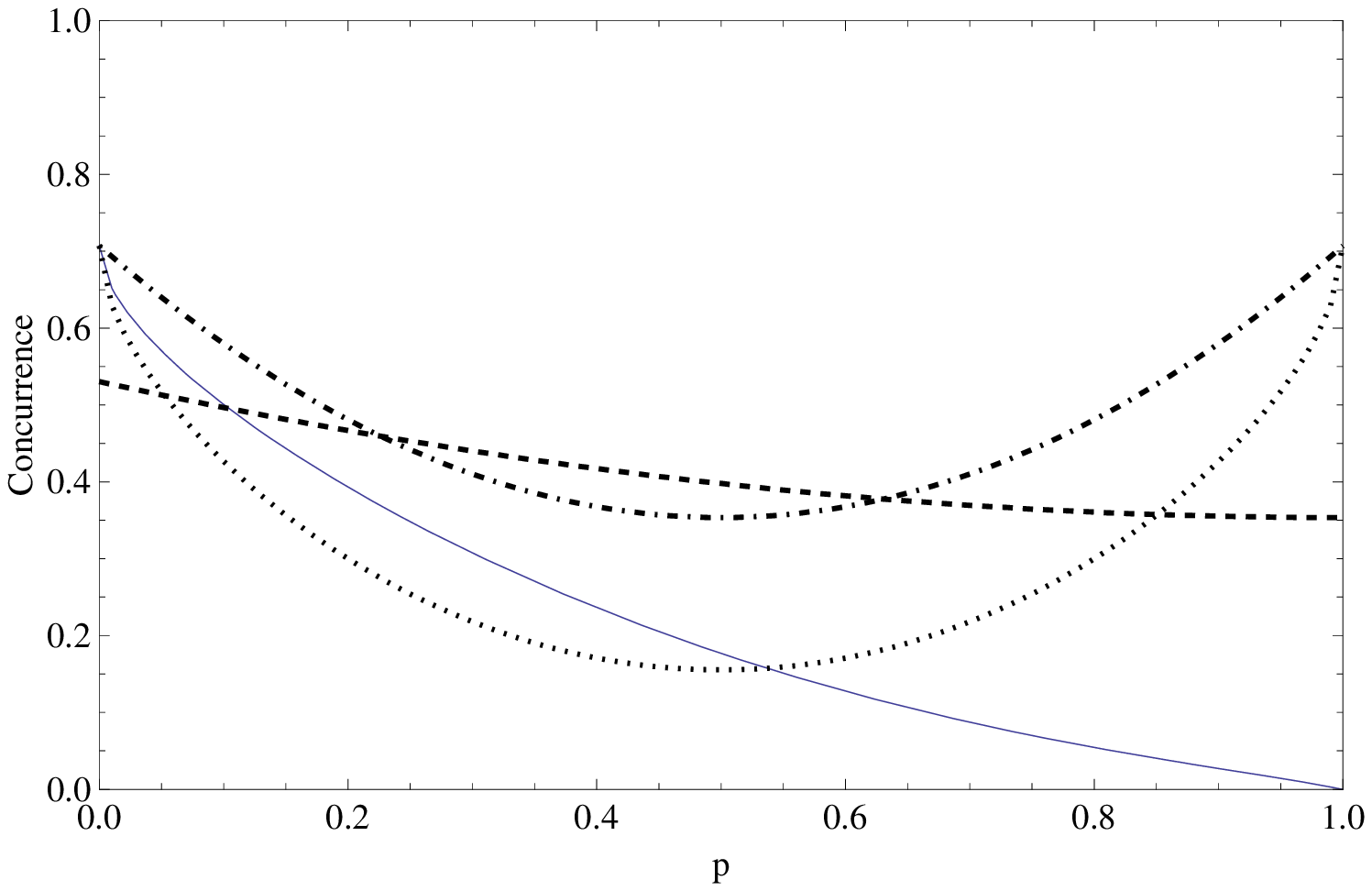} \\[0pt]
\end{center}
\caption{The concurrence is plotted as a function of decoherence parameter $%
p $ for $\protect\mu =0.5$\ and $r=\protect\pi /6$ for amplitude damping
(solid line), depolarizing (dashed line), bit-phase flip (dotted line) and
phase flip (dot dashed line) channels.}
\end{figure}
\begin{figure}[tbp]
\begin{center}
\vspace{-2cm} \includegraphics[scale=0.8]{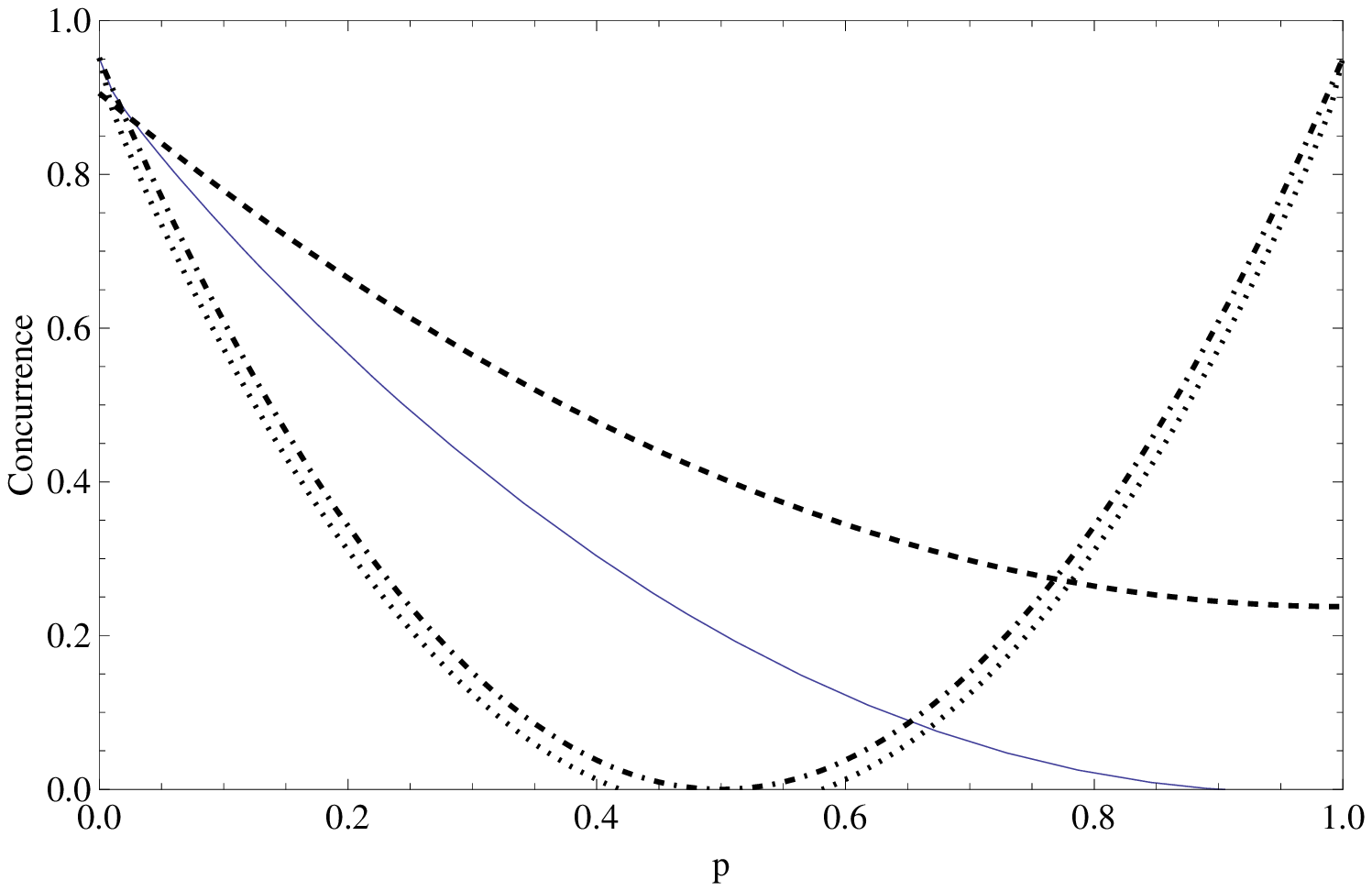} \\[0pt]
\end{center}
\caption{The concurrence is plotted as a function of decoherence parameter $%
p $ for $\protect\mu =0$\ and $r=\protect\pi /10$ for amplitude damping
(solid line), depolarizing (dashed line), bit-phase flip (dotted line) and
phase flip (dot dashed line) channels.}
\end{figure}
\begin{table}[tbh]
\caption{Single qubit Kraus operators for amplitude damping, depolarizing,
bit-phase flip and phase flip channels where $p$ represents the decoherence
parameter.}
\label{di-fit}$%
\begin{tabular}{|l|l|}
\hline
&  \\
$\text{Amplitude damping channel}$ & $A_{0}=\left[
\begin{array}{cc}
1 & 0 \\
0 & \sqrt{1-p}%
\end{array}%
\right] ,$ $A_{1}=\left[
\begin{array}{cc}
0 & \sqrt{p} \\
0 & 0%
\end{array}%
\right] $ \\ \hline
&  \\
$\text{Depolarizing channel}$ & $%
\begin{tabular}{l}
$A_{0}=\sqrt{1-\frac{3p}{4}I},\quad A_{1}=\sqrt{\frac{p}{4}}\sigma _{x}$ \\
$A_{2}=\sqrt{\frac{p}{4}}\sigma _{y},\quad \quad $\ $\ A_{3}=\sqrt{\frac{p}{4%
}}\sigma _{z}$%
\end{tabular}%
$ \\
&  \\ \hline
$\text{Bit-phase flip channel}$ & $A_{0}=\sqrt{1-p}I,\quad A_{1}=\sqrt{p}%
\sigma _{y}$ \\
&  \\ \hline
$\text{Phase flip channel}$ & $A_{0}=\sqrt{1-p}I,\quad A_{1}=\sqrt{p}\sigma
_{z}$ \\ \hline
\end{tabular}%
$%
\end{table}

\end{document}